\documentstyle[prb,preprint,aps]{revtex}

\begin{document}
\draft
\title{Electron correlations for ground state properties
of group IV semiconductors}

\author{Beate Paulus, Peter Fulde}
\address{Max-Planck-Institut f\"ur Physik komplexer Systeme,
Bayreuther Str. 40, D-01187 Dresden}
\author{Hermann Stoll}
\address{Institut f\"ur Theoretische Chemie, Universit\"at Stuttgart,
D-70550 Stuttgart}
\date{21st of October 1994}
\maketitle

\begin{abstract}
Valence energies for crystalline C, Si, Ge, and Sn with diamond structure
have been determined using an ab-initio approach based on information
from cluster calculations.
Correlation contributions, in particular, have been
evaluated in the coupled electron pair approximation (CEPA),
by means of increments obtained
for localized bond orbitals and for pairs and triples of such bonds.
Combining these results with corresponding Hartree-Fock (HF) data,
we recover about 95 \% of the experimental cohesive
energies.
Lattice constants  are overestimated
at the  HF level by about 1.5 \%;
correlation effects reduce these deviations to values
which are within the error bounds of this method.
A similar behavior is found for the bulk modulus: the HF values which
are significantly too high are reduced by correlation
effects to $\sim 97$\% of the experimental values.
\end{abstract}


\section{Introduction}

Several methods are presently available for performing {\it ab initio}
calculations for solids. Most frequently used is the density-functional
method, with a local-density approximation (LDA) for the exchange and
correlation contributions to the total energy \cite{lda}. This method
yields good results for solid-state properties, on average,
but a systematic improvement is difficult.
In the past few years it has become possible to determine
self-consistent-field (SCF)
calculations for solids using the exact non-local Hartree-Fock (HF)
exchange \cite{scf}.
While the results are often inferior to LDA  because
electron correlations are neglected, it has the advantage of yielding a
well-defined mean-field wave-function, which can be used as
a starting point for treating many-body effects.
In the Quantum Monte Carlo (QMC) approach \cite{qm}, one multiplies a mean
field function (Slater determinante) by a Jastrow factor
which explicitly introduces
inter-electronic coordinates.
Another approach of incorporating electron correlation is the local ansatz (LA)
\cite{la}, where local operators acting on the SCF wave-function are used to
admix suitable one- and two-particle excitations
to the mean-field ground state.
A method closely related to the ideas of the LA is the incremental expansion of
the correlation energy; here information on local excitations in clusters
is made use of,
which are accessible to an accurate quantum-chemical configuration
interaction (CI) treatment. This way
$\sim 85$ \% of the correlation
contribution to the cohesive energy $E_{coh}$ has been recovered for
diamond \cite{diamond} and
crystalline silicon \cite{silicon}.
In the present paper we extend the application of the method of increments to
two more systems
-- germanium and grey tin -- and to other properties beside $E_{coh}$:
we determine, for all of the group IV semiconductors,
the influence of  electron correlation on
the lattice constant $a$ and the bulk modulus $B$.

In order to discuss the influence of correlation, we need reliable HF values
for the ground state properties of these solids. We supplement the
results of the Torino group \cite{crysvalue} by  cluster
calculations which are described in Section 2.
In Section 3, we sketch the method of increments and
report on  computational details of the correlation treatment.
The results for  group IV
semiconductors are discussed and compared with experiment and
other calculations in
Section 4. Conclusions follow in Section 5.

\section{Hartree-Fock Calculations}

Obtaining accurate HF energies for solids is still a difficult computational
task. The Torino group of Pisani and co-workers developed a method which
enables one to perform SCF-LCAO (linear combination of atomic orbitals)
calculations for periodic systems
\cite{scf}. However, as a certain drawback of their program package
{\sc Crystal} \cite{crystal} convergency problems are encountered,
if one uses
Gaussian functions with small exponents as normally contained in
molecular basis sets; moreover, higher polarization functions than
$d$ are presently not available.

Another possibility for obtaining information on solids
are cluster calculations. They do not account for the infinite
extension of the solid, but one can use
standard quantum chemical program packages for achieving
high accuracy, without any restriction with regard to the basis set.
In this paper, we use a cluster method in conjunction with an energy
partitioning approach which was developed for diamond \cite{cluster} and is
here extended
to silicon, germanium and grey tin. The systems considered in this scheme
are fragments of the diamond lattice, with dangling bonds saturated by
hydrogen atoms.
More specifically, one chooses clusters with closed structure where
each of the X atoms (X=C, Si, Ge, Sn) has at least two neighbouring X atoms.
The smallest one which meets this condition is $\mbox{X}_6\mbox{H}_{12}$,
the biggest one which was accessible in our
calculations is $\mbox{X}_{35}\mbox{H}_{36}$. (All
clusters of the present SCF study are shown in Fig. 1.)
The X-X distances were taken from the solid ($r_{\mbox{CC}}$=1.544 \AA ,
$r_{\mbox{SiSi}}$=2.352 \AA , $r_{\mbox{GeGe}}$=2.460 \AA ,  $r_{\mbox{SnSn}}
$=2.810
\AA )
\cite{landolt1}, for the X-H bond lengths those of the corresponding
$\mbox{XH}_4$ molecule were used
($r_{\mbox{CH}}$=1.102 \AA , $r_{\mbox{SiH}}$=1.480 \AA ,
$r_{\mbox{GeH}}$=1.525 \AA , $r_{\mbox{SnH}}$=1.711 \AA ) \cite{landolt2}.
For carbon the correlation-consistent polarized
valence double-zeta (pvdz) basis set
(9s4p1d)/[3s2p1d] of
Dunning \cite{dunning} was selected.
For Si, Ge and Sn we employed  4-valence-electron pseudopotentials
simulating  the atomic cores, together with
the corresponding optimized basis sets of double-zeta quality (dz)
(4s4p)/[3s3p]
for the valence electrons \cite{pseudo}; we added one d-function
in each case whose exponent was optimized
in CI calculations for $\mbox{XH}_4$ and $\mbox{X}_2\mbox{H}
_6$ (Si: 0.40; Ge: 0.32; Sn: 0.23). For hydrogen we chose Dunning's
\cite{dunning} double zeta basis
(4s)/[2s], without polarization function.
We calculated the total HF energy $E_{\mbox{total}}$ for each cluster, using
the
direct SCF program package {\sc Turbomole} \cite{turbomol} of Ahlrichs
and co-workers. In order to obtain an estimate of that part of
$E_{\mbox{total}}$ which can be attributed to a 'solid-like' X atom of the
cluster,
surrounded only by X atoms, we
employed the energy partitioning method of Refs. \cite{cluster,epm1,epm2,epm3}:
$E_{\mbox{total}}$  is approximated by a sum of energy contributions of the
bare X atom, XH and ${\mbox{XH}}_2$ groups
\begin{equation}
E_{\mbox{total}}=n_{\mbox{X}}E_{\mbox{X}}+
n_{\mbox{XH}}E_{\mbox{XH}}+n_{\mbox{XH}_2}E_{\mbox{XH}_2}
\end{equation}
where $n_{\mbox{X}}$ ($n_{\mbox{XH}}$, $n_{\mbox{XH}_2}$) are the numbers
of atoms in the cluster with zero (one, two) neighbouring H atoms.
We determined  the quantities $E_{\mbox{X}}$, $E_{\mbox{XH}}$, and $E_{
\mbox{XH}_2}$
by adjustment to the SCF results $E_{\mbox{total}}$ of
the three
largest clusters ($\mbox{X}_{22}\mbox{H}_{28}$, $\mbox{X}_{26}\mbox{H}_{30}$
and $\mbox{X}_{35}\mbox{H}_{36}$). We checked the resulting values for the
group
energies using the total SCF energies of the  other clusters:
the quantity $\sigma$ in Table 1 is the (average) difference
of these energies to the values evaluated from the group contributions
according
to (1).
Finally, the HF cohesive energy of the solid per unit cell
was calculated from
\begin{equation}
E_{\mbox{coh}}=n_{\mbox{uc}}(E_{\mbox{X}}-E_{\mbox{atom}})
\end{equation}
where $E_{\mbox{atom}}$ is the SCF energy of the free atom determined at the
same level as $E_{\mbox{X}}$ (i.e. using the same basis set), and
$n_{\mbox{uc}}$
is the number of atoms per unit cell (2 for the diamond lattice).
The results for $E_{\mbox{coh}}$ of the group IV semiconductors are
listed in Table 1. The error $\sigma$ gives an estimate for
the finite size effect; it is
3 \% for Sn and smaller for the
other compounds.
Hartree-Fock results from literature \cite{crysvalue},
calculated  with {\sc Crystal},
are in good agreement (to $\sim 1\%$) with our values for diamond and
silicon (although smaller primitive basis sets were used in
Ref.\ \cite{crysvalue}), but there is a difference of 7 \% for Ge.

\section{Correlation-energy  increments}
\subsection{Formalism}

In the following we give a derivation of the
 incremental expansion for the correlation energy
which supplements the one given in Refs.\ \cite{diamond,silicon};
it is similar to the
one proposed in Ref.\ \cite{tomdr}. The present
derivation is more formal,
but it shows clearly that the method applies to infinite
solids and should not be considered as being merely related to
calculations for  finite
systems. We use thereby the CEPA-0 (see e.g. Ref.\ \cite{cepa}) which is
particularly suitable for our purpose.

We start from a Hamiltonian $H$ which can be divided into two parts
\begin{equation}
H=H_0+H_1.
\end{equation}
The ground state of $H_0$ is supposed to be known; in our case it is the
HF ground state with the corresponding wave-function $\Phi_0$. We define
a product of two operators $A$ and $B$ in the Liouville space as
follows:
\begin{equation}
(A|B)=\langle\Phi_0|A^{\dagger}B|\Phi_0\rangle ^{\mbox{c}}=\langle A^
{\dagger}B\rangle ^
{\mbox{c}}\label{cumm}
\end{equation}
The upper script c indicates that the cumulant of the
expectation value is taken, which is given by
\begin{eqnarray}
\langle A\rangle ^{\mbox{c}}&=&\langle A\rangle \\
\langle AB\rangle ^{\mbox{c}}&=&\langle AB\rangle -\langle A\rangle \langle B
\rangle \\
&&\hbox{etc.} \nonumber
\end{eqnarray}
For further details see Ref.
\cite{cummulants}. By using (\ref{cumm}) we may write the exact
ground-state energy $E$ in the following way:
\begin{equation}
E=(H|\Omega )=E_{\mbox{HF}}+(H_1|\Omega ),\label{energy}
\end{equation}
where $\Omega$ plays the role of the wave operator which describes the
transformation from the HF ground state to the exact ground state.

If we have a solid with well-defined bonds, we can express the HF
ground state $\Phi_0$ in terms of localized orbitals and label
those orbitals by a bond index $i$.
We define operators $A_i$,
where i should being consider as a compact index which includes the
bond i as well as the one and two particle
excitations of bond $i$, and $A_{ij}$, which
describes the two particle excitations
where one excitation is out of bond $i$ while the
other is out of bond $j$. Within the restricted operator subspace
spanned by $A_i$ and $A_{ij}$
we make for $\Omega$ an ansatz of the form
\begin{equation}
|\Omega )=|1+\sum_i n_i A_i + \sum_{ij \atop i \not= j}n_{ij}A_{ij}).
\label{omega}
\end{equation}
This choice represents the coupled electron pair approximation at level
zero (CEPA-0) \cite{fulde}.
The parameters $n_i$ and $n_{ij}$ are determined
from the set of equations $(A_i|H\Omega)=0$ and $(A_{ij}|H\Omega)=0$
\cite{tom}.
With (\ref{omega}) this implies
\begin{eqnarray}
0&=&(A_k|H)+\sum_i n_i (A_k|HA_i)+\sum_{ij \atop i \not= j}n_{ij}
(A_k|HA_{ij}) \nonumber \\
0&=&(A_{kl}|H)+\sum_i n_i (A_{kl}|HA_i)+\sum_{ij \atop i \not= j}n_{ij}
(A_{kl}|HA_{ij}). \label{set}
\end{eqnarray}
The method of increments provides a scheme,
in which this set of equations and hence the correlation energy is
evaluated in a hierarchical order.
\begin{itemize}
\item[a)]
First all electrons are kept frozen except for the ones e.g. in bond $i$.
The operators $A_i$ describe the corresponding excitations of these two
electrons and Eq. (\ref{set}) reduce to
\begin{equation}
0=(A_i|H)+n_i^{(1)}(A_i|HA_i).
\end{equation}
Within this approximation the $n_i^{(1)}$ are independent of each other
and the correlation energy becomes
\begin{equation}
E_{\mbox{corr}}^{(1)}=\sum_i \epsilon_i \label{corren}
\end{equation}
with
\begin{equation}
\epsilon_i=n_i^{(1)}(H_1|A_i).
\end{equation}
\item[b)]
In the next step we correlate the electrons in two bonds,
e.g. $i$ and $j$.
The corresponding $n^{(2)}$ parameters are determined from the
coupled equations
\begin{eqnarray}
0&=&(A_i|H)+n_i^{(2)}(A_i|HA_i)+n_j^{(2)}(A_i|HA_j)+n_{ij}^{(2)}(A_i|HA_{ij})
\nonumber\\
0&=&(A_j|H)+n_i^{(2)}(A_j|HA_i)+n_j^{(2)}(A_j|HA_j)+n_{ij}^{(2)}(A_j|HA_{ij})
\nonumber\\
0&=&(A_{ij}|H)+n_i^{(2)}(A_{ij}|HA_i)+n_j^{(2)}(A_{ij}|HA_j)+n_{ij}^{(2)}
(A_{ij}|HA_{ij})
\end{eqnarray}
Again, the increments $\delta n_i=n_i^{(2)}-n_i^{(1)}$ and
$\delta n_j=n_j^{(2)}-n_j^{(1)}$ are treated as independent of each other in
this approximation, and we have
\begin{equation}
E_{\mbox{corr}}^{(2)}=\sum_i \epsilon_i
+\frac{1}{2}\sum_{ij \atop i \not= j} \Delta \epsilon_{ij}
\end{equation}
where
\begin{equation}
\Delta\epsilon_{ij}=\epsilon_{ij}-(\epsilon_i+\epsilon_j).
\end{equation}
and
\begin{equation}
\epsilon_{ij}=(H_1|n_i^{(2)}A_i+n_j^{(2)}A_j+n_{ij}^{(2)}A_{ij}).
\end{equation}
\item[c)]
Analogously we calculate the three bond energy increment, which is defined as
\begin{equation}
\Delta \epsilon_{ijk}= \epsilon_{ijk}
-(\Delta \epsilon_i + \Delta \epsilon_j + \Delta \epsilon_k)
-(\Delta \epsilon_{ij}+\Delta \epsilon_{jk}+ \Delta \epsilon_{ik}).
\end{equation}
The correlation energy $\epsilon_{ijk}$ is that obtained when all electrons
are kept frozen except those in bond $i$,$j$ and $k$.
Again, the increments $\Delta \epsilon_{ijk}$ are treated as being
independent of each other.
\end{itemize}
The total correlation energy within this approximation is the sum of all
increments.
\begin{equation}
E_{\mbox{corr}}^{\mbox{solid}} = \sum_i  \epsilon_i +
\frac{1}{2}\sum_{ij \atop i \not= j} \Delta \epsilon_{ij}+
\frac{1}{6}\sum_{ijk
\atop i \not= j \not= k}
\Delta \epsilon_{ijk} + ...
\end{equation}
It is obvious that by calculating higher and higher increments
the exact correlation energy within CEPA-0 is obtained.

The method just described is only useful if the incremental
expansion is well convergent, i.e. if increments up to, say, triples are
sufficient, and if increments
become rapidly small with
increasing distance between localized orbitals.
These conditions were shown to be well met in the case of diamond and silicon
\cite{diamond,silicon}, but have to be checked again for germanium and
grey tin here.
Ideally the increments should be local entities not sensitive to the
surroundings. We use this property to calculate the energy increments
in finite clusters.
If they can be proven to be well transferable even between
finite clusters, such cluster calculations may be extrapolated
to the corresponding solid state quantities.

\subsection{Computational Details}

In this section we give computational details characterizing our
{\it ab initio} calculations for the correlation energies
of group IV semiconductors with diamond structure.

First we select suitable fragments of the diamond lattice, and, as in the
SCF calculations, we saturate the dangling bonds with hydrogen. We have to
take much smaller clusters,
however, for the correlation treatment (see Fig.\ 2) than in the SCF case,
because CI calculations for clusters as large as
$\mbox{X}_{35}\mbox{H}_{36}$ would be prohibitive.

As a second step we perform, for each cluster, a standard SCF calculation
(using the program package {\sc Molpro} \cite{molpro})
and localize the bond orbitals, applying
the Foster-Boys
criterion \cite{foster}, within the occupied valence space in $\mbox{C}_1$
symmetry. Following that we calculate CEPA-0  energy increments
by successively correlating more and more of the localized X-X bond orbitals
(LMO) as described in the previous subsection.

Two different basis sets are used: Basis set A is the same as in
the SCF calculations. An extended basis set B has been
generated by
replacing the single $d$ function of basis set A by a $2d1f$ polarization
set with an optimized $f$ exponent (C: d: 1.0970, 0.318; f: 0.76; Si:
d: 0.8, 0.23; f: 0.35; Ge: d: 0.6, 0.15; f: 0.43; Sn: d: 0.45, 0.12; f: 0.30).
In order to check the quality of the basis sets we performed
test calculations for  the one-bond increment
$\Delta \epsilon_i$ of the $\mbox{Ge}_2\mbox{H}_6$ cluster (Table 2).
One sees that the $sp$ basis set is well chosen; enlarging the $d$
space and adding
an additional $f$ function
yields an enhancement of about 25\% , supplying
a polarisation set 3d2f1g
\cite{Steinbrenner} adds another 20\% .

Next we have to check the transferability and the convergence
of the increments. We will discuss them for Sn since this is the most
critical case.
The correlation energy increments obtained for Sn are listed in Table 3
together with the weight factors appropriate for the solid.
We observe the following:
\begin{itemize}
\item[a)] The convergence of the incremental expansion is quite rapid
both with respect to the number of the bonds involved (e.g. cluster 4:
$\epsilon_2 \approx 2.5 \Delta \epsilon_{23} \approx 300 \Delta \epsilon_{123}
 \approx 1500 \Delta \epsilon_{1234}$) and with respect
to the distance between the bonds
(e.g. cluster 4: $\Delta \epsilon_{12} \approx 10 \Delta \epsilon_{13}
\approx 50 \Delta \epsilon_{14}$).
In view of these findings, we restrict the energy increments up to the third
order for adjacent bonds and up to the second order for next nearest through
third nearest neighbours.
\item[b)] The transferability of the increments is reasonably good.
There are changes of $1\cdot 10^{-3}$ a.u. for one-bond increments between
the smallest ($\mbox{Sn}_2\mbox{H}_6$) and the largest
($\mbox{Sn}_6\mbox{H}_{14}
$) clusters considered. For the most important two-bond increment the
difference
between $\mbox{Sn}_3\mbox{H}_8$ and $\mbox{Sn}_6\mbox{H}_{14}$  is about
$1\cdot 10^{-4}$ a.u.. As these changes are of different sign, the
total effect on the cohesive energy of the solid is not larger
than $3\cdot 10^{-3}$ a.u. per unit
cell.
\end{itemize}

Finally, in order to test the quality of the CEPA-0 results, we performed
 calculations for $\mbox{Ge}_3
\mbox{H}_8$ at different levels of correlation treatment (Table 4).
We tried CEPA-1 and CEPA-2 as well as the coupled cluster method
with single and double  excitations (CCSD) and even with
additional triple excitations
included in a perturbative way
(CCSD(T)). The effect on the one-bond increment is of the order of
$3\cdot 10^{-4}$
a.u., and $\sim 2\cdot 10^{-4}$ a.u.\ for the two-bond increment. Thus,
the errors are
of the same order of magnitude as those caused by lack of transferability.

It is clear, from these remarks, that the accuracy of our method has its
limitations, in practical applications.
On the other hand, there are formal arguments why errors due to the
truncation of the incremental expansion should be small: higher than two-bond
increments involve
triple excitations which do not directly couple to the Hartree-Fock
ground state; two-bond
increments between localized orbitals at large distances are of the
van-der-Waals type, with a rapid decrease as $\frac{1}{r^6}$.
Moreover, there are means for numerically controlling these errors:
the $\frac{1}{r^6}$ law just mentioned, e.g.,
lends itself to an approximate summation
of neglected two-bond increments; energy variations between different finite
clusters allow for an easy estimate of
transferability defects of individual local increments.

Summarizing these considerations, one may conclude  that
the greatest remaining error of the results, to be discussed in the next
section, is due to
limitation of the one-particle basis sets.

\section{Results and discussion}

Applying the method of increments as described in the preceding section,
we have determined
correlation contributions to cohesive energies, lattice constants, and
bulk moduli
for all of the  group IV
semiconductors. The increments were always taken from the largest
possible cluster (cf.\ Table 3 and Fig.\ 2).

The
correlation contributions to the cohesive energies were obtained as
$ E_{\mbox{coh}}^{\mbox{corr}}=E_{\mbox{solid}}^{\mbox{corr}}-2E_{\mbox{atom}}
^{\mbox{corr}}$ per
unit cell of the diamond lattice. The results for the two different
basis sets are shown in Table 5. For basis set A we obtain about
65 \% of the `experimental' correlation energies (defined here as the
differences between the experimental cohesive energies and the
corresponding HF values
of Sect.\ 2).
The larger basis set B yields  a  substantial improvement,
to about 85 \% . Combining the HF results with
the correlation contributions we recover about 95 \% of the experimental
cohesive energies \cite{expcoh}.
(The experimental values in Table 5 has been corrected for the phonon
zero point energies $\frac{9}{8} k_{\mbox{B}} \Theta_{\mbox{D}}$
derived from the Debye model
\cite{debye} (C: 1860 K; Si: 625 K; Ge: 360 K; Sn: 260 K) \cite{detem}.)

For comparison, we have also listed in Table 5
results from the literature which have been obtained with other methods.
The Local Ansatz (LA) which also uses a  CEPA-0 scheme yields smaller
cohesive energies for all compounds.
\cite{roland} (The LA values in Table 5 have been obtained by adding
the
correlation contributions to $E_{\mbox{coh}}$ calculated in Ref \cite{roland}
to the SCF values of the present work (Sect. 2).)
Our basis set A is comparable with the basis
used in LA, but still the correlation contributions differ by
about 1 eV for Si, Ge and Sn.
LDA \cite{qm,ldaresults}
overestimates the cohesive energies by $\sim$15 \%.
The QMC result \cite{qm} for diamond
is excellent and also very good for silicon (with an accuracy of 4\%);
note, however,
that in both cases the
HF cohesive energy of Ref.\ \cite{qm} is  lower by $\sim 1$ eV than that of
Ref.\ \cite{crysvalue}
and of this work.

In the next step we evaluate lattice constants (Table 6). They  have
been determined by varying the X-X distances both in the HF
and in the CEPA-0 calculations.
More specifically, to obtain the HF lattice constant we varied
all X-X distances of the
largest cluster $\mbox{X}_{35}\mbox{H}_{36}$ and minimized the SCF energy
with respect to the interatomic distance.
At the HF level, the lattice constant is larger than in
the experiment, except for
diamond. Our results are in fair agreement here with the {\sc Crystal} SCF
calculations by Causa and Zupan \cite{crysvalue}:
the deviations are $\sim 0.5 $\%
for C and Si, but $\sim 1$\% for Ge, see Table 6.
By using pseudopotentials for the core electrons in the present work,
any effects of the latter beyond the frozen-core approximation are excluded
from the outset. However,
calculations for small molecules show
that core polarization is very important for bond lengths of
Ge and Sn compounds \cite{cp}.
We simulated, therefore, this effect using a core polarization
potential (CPP) \cite{cpp} and studied  the influence on
the X-X distance for different clusters. We find that the
changes are nearly independent of the
cluster size. By transfering these changes to the solid we obtain the
results listed in row b) of Table 6. The influence of core polarization
is seen to significantly increase within the group
Si, Ge and Sn.
Valence correlations affect the lattice constant in two different
ways.
Correlation contributions calculated with a minimal basis set
(`inter-atomic correlations') enlarge the
lattice constant. The increase for diamond is mostly due to
them. Intra-atomic correlations, on the other hand,
decrease the lattice constant; extended
basis sets with many polarization
functions are needed to describe them accurately.
The total influence of valence correlation seems to become less
important when going from Si to Ge and Sn.
The final values are within the error bounds of the
pseudopotential and CPP ($\approx 0.02$ \AA),
and within the error of the limited basis set which
can be roughly estimated from the difference between basis sets
A and B to be $\leq 0.02$ \AA.
The LA overestimates the lattice constant, since  with the small basis set
used in Ref.\ \cite{roland} the intra-atomic correlations could not be
described very well.
The LDA yields lattice constants which are systematically too small
by about 1\%.
The QMC result is again excellent for diamond, slightly less good for silicon,
but still reaching the experimental value within their error bounds.

As a last property, we consider the bulk modulus $B=V\frac{\partial^2 E}
{\partial V^2}$, which describes the response of the solid to a
homogeneous pressure. For the diamond structure
one can easily derive the following expression
\begin{equation}
B=\left( \frac{4}{9a}\frac{\partial^2}{\partial a^2} - \frac{8}{9a^2}
\frac{\partial}{\partial a} \right) E_{\mbox{coh}} (a).
\label{bulk}
\end{equation}
We evaluate the bulk modulus at the experimental lattice constant $a$,
so that the second term in (\ref{bulk}) is small but not zero.
We obtain the HF bulk modulus again from the largest possible cluster.
Compared with  experiment \cite{landolt1} (see Table 7) all values are too
high.
The core-polarization effect has been taken
from smaller clusters, which causes an uncertainty of about 3 \%, but the
total effect of the CPP is much greater so that one can justify this
approximation: for Sn, it reduces B by 14 \%.
Valence correlation
reduces the bulk modulus, too, especially for Ge.
Overall, we obtain results for the bulk modulus which are a little
too small but still within few percents of the experimental data.

\section{Conclusions}

We have determined ground state properties (cohesive energy, lattice
constant and bulk modulus) of group IV semiconductors, both at
the HF and the CEPA-0 levels.
The HF results have been obtained using a  cluster method and an
energy-partitioning ansatz which works
well for solids with covalent bonds as those considered here.
Electronic correlations at the CEPA-0 level are described with the method
of local intra- and inter-bond increments,
which allows for a systematic improvement of accuracy
towards the fully correlated solid-state limit.
The results show that this method  works well for all
solids with diamond structure,
with the transferability only slightly deteriorating
down the fourth column of the Periodic Table.
Not only the cohesive energy E$_{\mbox{coh}}$, but also the lattice constant
$a$,
and the bulk modulus $B$, have been
calculated with quantum chemical accuracy, to about 5\% for E$_{coh}$, 0.5\%
for $a$, and 3\% for $B$.
Work is underway in our laboratory to apply the methods presented in
this paper to solids with
zinkblende
structure, such as GaAs.

\section{Acknowledgments}

We are grateful to Prof.\ H.-J.\ Werner, Stuttgart,
and to Prof.\ R.\ Ahlrichs, Karlsruhe, for providing their programs
{\sc Molpro} and {\sc Turbomole} respectively. We also thank
A. Nickla\ss , Stuttgart, for implementing CPP
routines into MOLPRO.
One of us (B.\ P.) appreciated several useful discussions with Dr.\ R.\
Pardon and Dr.\ T.\ Schork.

\begin{table}[h]
\caption{Hartree-Fock cohesive
energies per unit cell (in a.u.), with average deviations $\sigma$ in
parentheses (cf. text)}
\begin{tabular}{|c||c|c|c|c|}
&C&Si&Ge&Sn\\
\hline\hline
this work&-0.3947&-0.2273&-0.1560&-0.1323\\
&(0.0036)&(0.0032)&(0.0034)&(0.0044)\\
\hline
Ref.\ \cite{crysvalue}&-0.3984&-0.2253&-0.1672&---
\end{tabular}
\end{table}
\begin{table}
\caption{Test calculations for the Ge basis set:
the SCF energy, E$_{\mbox{SCF}}$, and the one-bond CEPA-0 correlation-energy
increment,
$\Delta \epsilon_i$,
for $\mbox{Ge}_2\mbox{H}_6$
(in a.u.)}
\begin{tabular}{|lll|c|c|}
&basis set&&$E_{\mbox{SCF}}$&$\Delta \epsilon_i$\\
\hline
A&(4s4p)/[3s3p]& 1d&-10.831779&-0.01845\\
B&(4s4p)/[3s3p]& 2d1f&-10.836493&-0.02265\\
&(4s4p)&2d1f&-10.836633&-0.02301\\
&(6s6p)&2d1f&-10.838324&-0.02317\\
&(6s6p)&3d2f1g&-10.841788&-0.02750
\end{tabular}
\end{table}

\begin{table}
\caption{Correlation-energy increments for Sn (in a.u),
determined at the
CEPA-0 level using basis set A. For the numbering of the
source clusters and bonds involved,
see Fig. 2.}
\begin{tabular}{|l|l|r|l|}
&Source cluster/&Increment&Weight factor\\
&bond orbitals&&for the solid\\
\hline\hline
$ \epsilon_i$&1/1& -0.018526&4\\
&2/1&-0.018257&\\
&4/2&-0.017966&\\
&8/1&-0.017476&\\
\hline
$\Delta \epsilon_{ij}$&2/1,2&-0.006794&12\\
&4/2,3&-0.006738&\\
&8/1,2&-0.006905&\\
&4/1,3&-0.000701&12\\
&8/2,5&-0.000715&\\
&7/1,3&-0.000575&24\\
&8/2,4&-0.000553&\\
&4/1,4&-0.000142&12\\
&5/1,4&-0.000104&48\\
&6/1,4&-0.000111&24\\
&7/1,4&-0.000194&12\\
\hline
$\Delta \epsilon_{ijk}$&8/1,2,3& 0.000475&8\\
&8/1,2,5&-0.000038&12\\
&4/1,2,3&-0.000055&\\
&8/1,2,4& 0.000055&24\\
\hline
$\Delta \epsilon_{ijkl}$&4/1,2,3,4&-0.000011&
\end{tabular}
\end{table}
\begin{table}
\caption{Test calculations for the one-bond increment
$\epsilon_i$ and the two-bond increment $\Delta \epsilon_{ij}$ between
nearest neighbours, at various levels of correlation treatment,
for $\mbox{Ge}_3\mbox{H}_8$
(a.u.)}
\begin{tabular}{|l|c|c|}
&$ \epsilon_i$&$\Delta \epsilon_{ij}$\\
\hline\hline
CEPA-0&-0.01824&-0.00651\\
CEPA-1&-0.01798&-0.00631\\
CEPA-2&-0.01798&-0.00650\\
CCSD&-0.01798&-0.00613\\
CCSD(T)&-0.01798&-0.00645
\end{tabular}
\end{table}

\begin{table}
\caption{Cohesive energies per unit cell (in a.u.);
deviations from experimental values (in percent) are given in parentheses}
\begin{tabular}{|c||c|c|c|c|}
&C&Si&Ge&Sn\\
\hline\hline
this work, basis set A &-0.5077&-0.2996&-0.2346&-0.1979\\
&(92 \% )&(87 \% )&(82 \%)&(86 \%)\\
\hline
this work, basis set B&-0.5276&-0.3248&-0.2565&-0.2230\\
&(96 \% )&(94 \% )&(90 \%)&(97 \%)\\
\hline
 exp. (see \cite{expcoh})&-0.555&-0.345&-0.285&-0.229\\
\hline
 LA (see\cite{roland})&-0.475&-0.259&-0.192&-0.164\\
\hline
LDA (see\cite{qm,ldaresults})&-0.634&-0.389&-0.333&---\\
\hline
QMC  (see\cite{qm})&-0.5475&-0.3587&---&---\\
\end{tabular}
\end{table}
\begin{table}
\caption{Lattice constants in \AA ngstr\"om
a) this work, SCF level, basis set A; b) this work, core polarization included,
basis set A; c) this work, valence correlation included, basis set B
-- in
comparison to experimental and other theoretical results.
Deviations from experimental values are given in parentheses}
\begin{tabular}{|c||c|c|c|c|}
&C&Si&Ge&Sn\\
\hline\hline
a)& 3.5590&5.4993&5.7516&6.6001\\
&(-0.2 \%)&(+1.2 \% )&(+1.7\% )&(+1.7\% )\\
\hline
b)&---&5.4662&5.6653&6.4549\\
&&(+0.6 \% )&(+0.2\% )&(-0.5\% )\\
\hline
c)& 3.5833&5.4256&5.6413&6.4443\\
&(+0.5 \%)&(-0.1 \% )&(-0.3\% )&(-0.7\% )\\
\hline
 exp. (see \cite{landolt1})& 3.5657&5.4317&5.6575&6.4892\\
\hline
 HF (see\cite{crysvalue})& 3.58&5.49&5.81&---\\
\hline
 LA (see\cite{roland})& 3.601&5.488&5.760&6.538\\
\hline
 LDA (see\cite{ldaresults})& 3.53&5.38&5.57&---\\
\hline
 QMC (see\cite{qm})& 3.543&5.404&---&---
\end{tabular}
\end{table}
\begin{table}
\caption{Bulk moduli
(in Mbar)
a) this work, SCF level, basis set A; b) this work, core polarization included,
basis set A; c) this work, valence correlation included, basis set B
-- in
comparison to experimental and other theoretical results.
Deviations from experimental values are given in parentheses}
\begin{tabular}{|c||c|c|c|c|}
&C&Si&Ge&Sn\\
\hline\hline
a)&4.815&1.038&0.961&0.638\\
&(+9 \%)&(+5 \% )&(+31\% )&(+20\% )\\
\hline
b)&---&1.009&0.889&0.562\\
&&(+2 \% )&(+21\% )&(+6\% )\\
\hline
c)&4.196&0.979&0.711&0.510\\
&(-5 \%)&(-1 \% )&(-3\% )&(-4\% )\\
\hline
 exp (see \cite{landolt1})&4.42&0.99&0.734&0.531\\
\hline
 HF (see \cite{crysvalue})&4.80&1.10&0.85&---\\
\hline
 LA (see\cite{roland})&4.332&1.001&0.774&0.509\\
\hline
 LDA (see\cite{ldaresults})&4.90&0.97&0.75&---\\
\hline
 QMC (see \cite{qm})&4.205&1.081&---&---
\end{tabular}
\end{table}

\begin{figure}
\caption{X$_n$H$_m$  clusters treated at the  SCF level
 (H-atoms are not drawn).}
\end{figure}
\begin{figure}
\caption{The X-skeletons of the clusters treated at the CEPA-0 level
(big numbers designate clusters,
small numbers the bonds in each cluster).}
\end{figure}

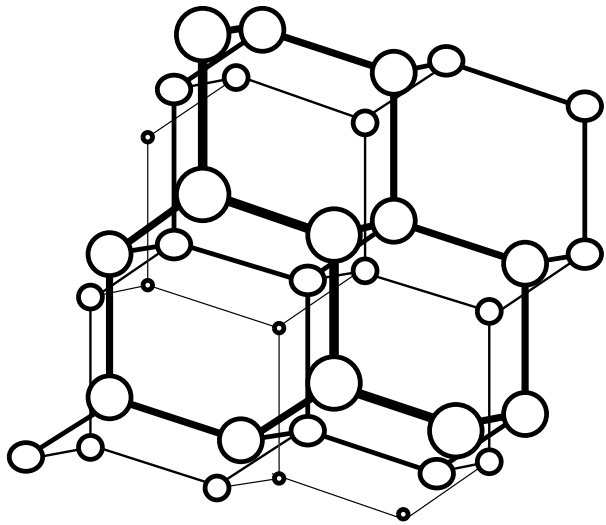
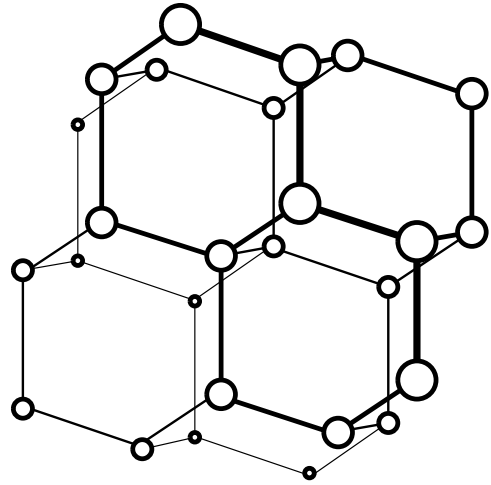
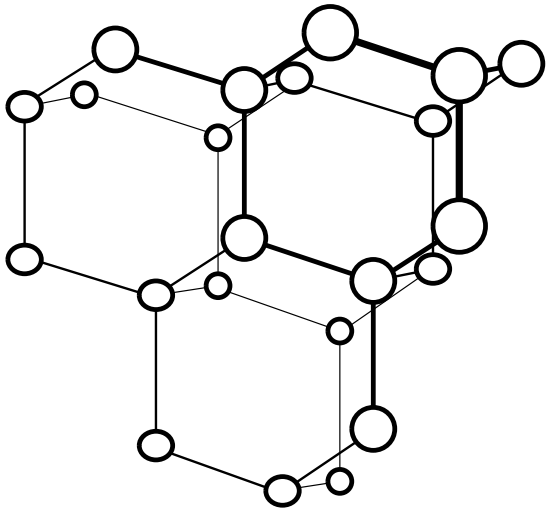
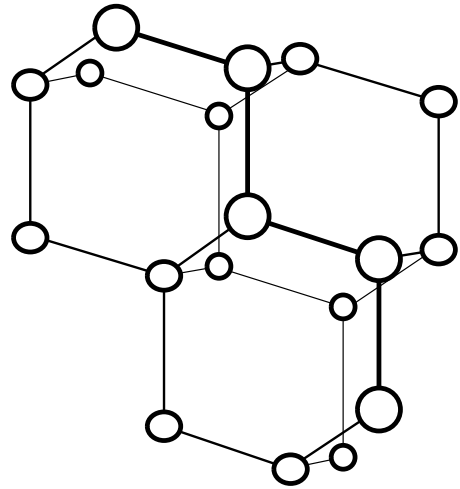
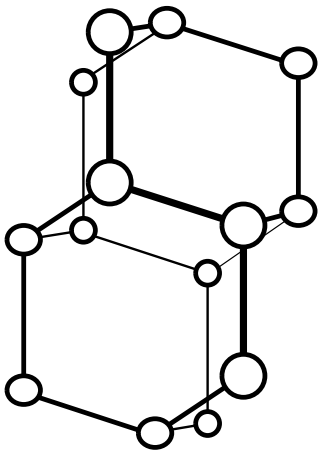
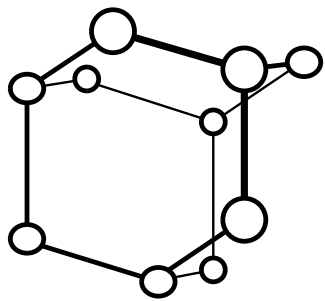
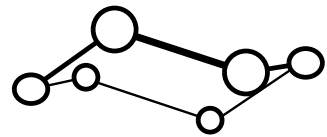
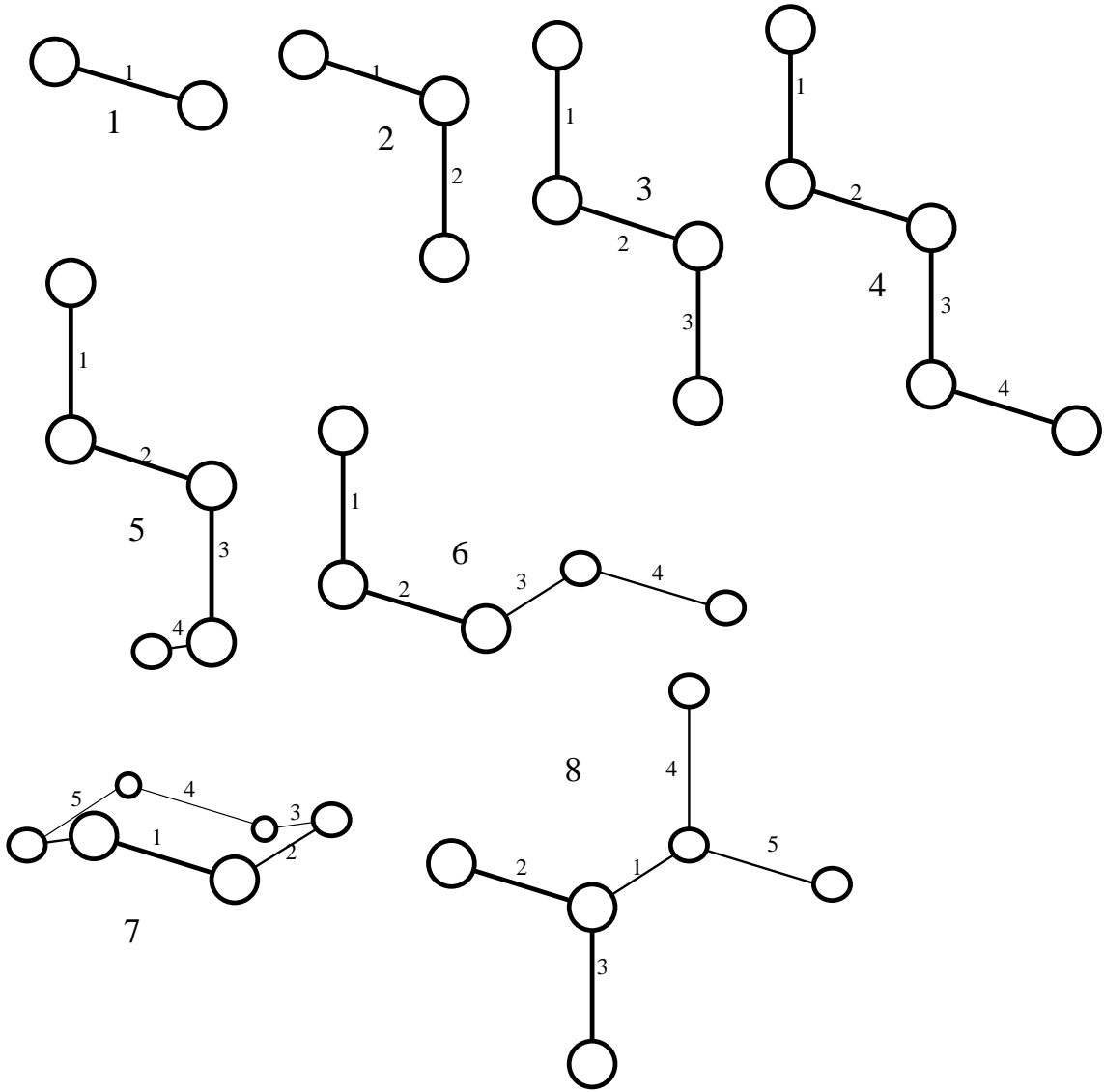
\end{document}